\documentclass[aps, prb, showpacs,amsmath,amssymb, preprint, floatfix, superscriptaddress]{revtex4}
\usepackage{graphicx}
\usepackage{bm}
\begin{document}

\title{High Pressure Structure and Decomposition of MoS$_{2}$}

\author{Oto Kohul\'{a}k}
\affiliation{Department of Experimental Physics, Comenius University, 
Mlynsk\'{a} Dolina F2, 842 48 Bratislava, Slovakia}

\author{Roman Marto\v{n}\'{a}k}
\email{martonak@fmph.uniba.sk}
\affiliation{Department of Experimental Physics, Comenius University, 
Mlynsk\'{a} Dolina F2, 842 48 Bratislava, Slovakia}

\author{Erio Tosatti}
\affiliation{International School for Advanced Studies (SISSA) and CNR-IOM Democritos, Via Bonomea 265, I-34136 Trieste, Italy}
\affiliation{The Abdus Salam International Centre for Theoretical Physics
  (ICTP), Strada Costiera 11, I-34151 Trieste, Italy}

\pacs{61.50.Ks, 71.30.+h, 72.80.Ga, 74.10.+v}

\date{\today}

\begin{abstract}
  The high pressure structural and electronic evolution of bulk
  MoS$_2$, an important transition metal layered dichalchogenide, is
  currently under active investigation. Recent theoretical and
  experimental work predicted and verified a 2H$_c \to$ 2H$_a$ layer
  sliding structural transition at 20 GPa and a band overlap
  semiconductor-semimetal transition in the same pressure range.  The
  2H$_a$ structure is known to persist up to pressure of 81 GPa but
  properties at higher pressures remain experimentally unknown. Here
  we predict, with a reliable first-principles evolutionary search,
  that major structural transformations should take place in
  equilibrium at higher pressures near 130-140 GPa. The main motif is
  a decomposition into MoS + S, also heralded in a small bimolecular
  cell by the appearance of a metastable non-layered metallic MoS$_2$
  structure with space group \textit{P4/mmm}.  Unlike semimetallic
  2H$_a$-MoS$_2$, both this phase and sulphur in the fully phase
  separated system are fully metallic and superconducting with higher
  critical temperatures than alkali-intercalated MoS$_2$.
\end{abstract}

\maketitle

MoS$_2$ is layered semiconductor, long known as a solid
lubricant,~\cite{lubricant} of more recent importance for its
applications and, in exfoliated form, as an electronic
material.\cite{PhysRevLett.105.136805,layer_transistor} It is also
relevant for superconductivity as a parent material that can be
metallized by alkali doping~\cite{somoano} or by surface field
doping~\cite{Taniguchi2012, Ye}.  In pursuit of superconductivity, it
is currently studied at ultra high pressures, where it was shown
theoretically\cite{Hromadova2013} and
experimentally\cite{PhysRevLett.113.036802} to undergo, besides a
layer sliding structural transition at 20 GPa from the 2H$_c$ to the
2H$_a$ structure, a band overlap metallization at 30-40 GPa
\cite{Hromadova2013, Nayak2014, PhysRevLett.113.036802}. The 2H$_a$
structure was experimentally shown to persist at least up to 81 GPa
[Ref.\cite{PhysRevLett.113.036802}] but the further evolution of
structure and electronic properties remain experimentally unclear at
higher pressures. Were the 2H$_a$ structure to persist further in that
regime, one might anticipate a relatively weak metallicity with modest
superconducting tendencies.~\cite{Hromadova2013, Profeta} There could,
however, be other structures coming up at ultra-high pressures,
possibly with a stronger metallicity and superconductivity.
Alternatively, the stability of MoS$_2$ as a compound might be
altogether lost giving rise, under sufficient pressure, to
decomposition and phase separation into products with different
stoichiometries and different properties.

We present here the results of first principle simulations which
predict that at thermodynamic equilibrium 2H$_a$-MoS$_2$ should become
unstable against both possibilities, phase transformation or
decomposition, when pressure reaches values between 130 and 140 GPa.
State-of-the-art density functional theory (DFT) calculations,
particularly accurate at higher pressure, can reliably explore the
equilibrium phase diagram of MoS$_2$ in this regime once they are
coupled with a suitable structure search algorithm.

Using the genetic search algorithm as implemented in the Xtalopt
package~\cite{Xtalopt} we performed a search of the lowest enthalpy
T=0 crystal structure of MoS$_2$ at 80, 120 and 150 GPa using
supercells with 6, 9 and 12 atoms, corresponding to 2, 3 and 4 formula
units. Using the PBE \cite{pbe} XC potential, we employed for the 6
and 9 atoms/cell the Quantum Espresso ~\cite{QE} code with a PW cutoff
of 60 Ry and a 4x4x4 k-point grid, and for the 12 atoms/cell the VASP
code ~\cite{VASP} with 450 eV cutoff and automatic k-point mesh
generation of length 30\footnote{see the VASP manual}.

\begin{figure}[h]
\centering
\includegraphics[width=0.95\textwidth]{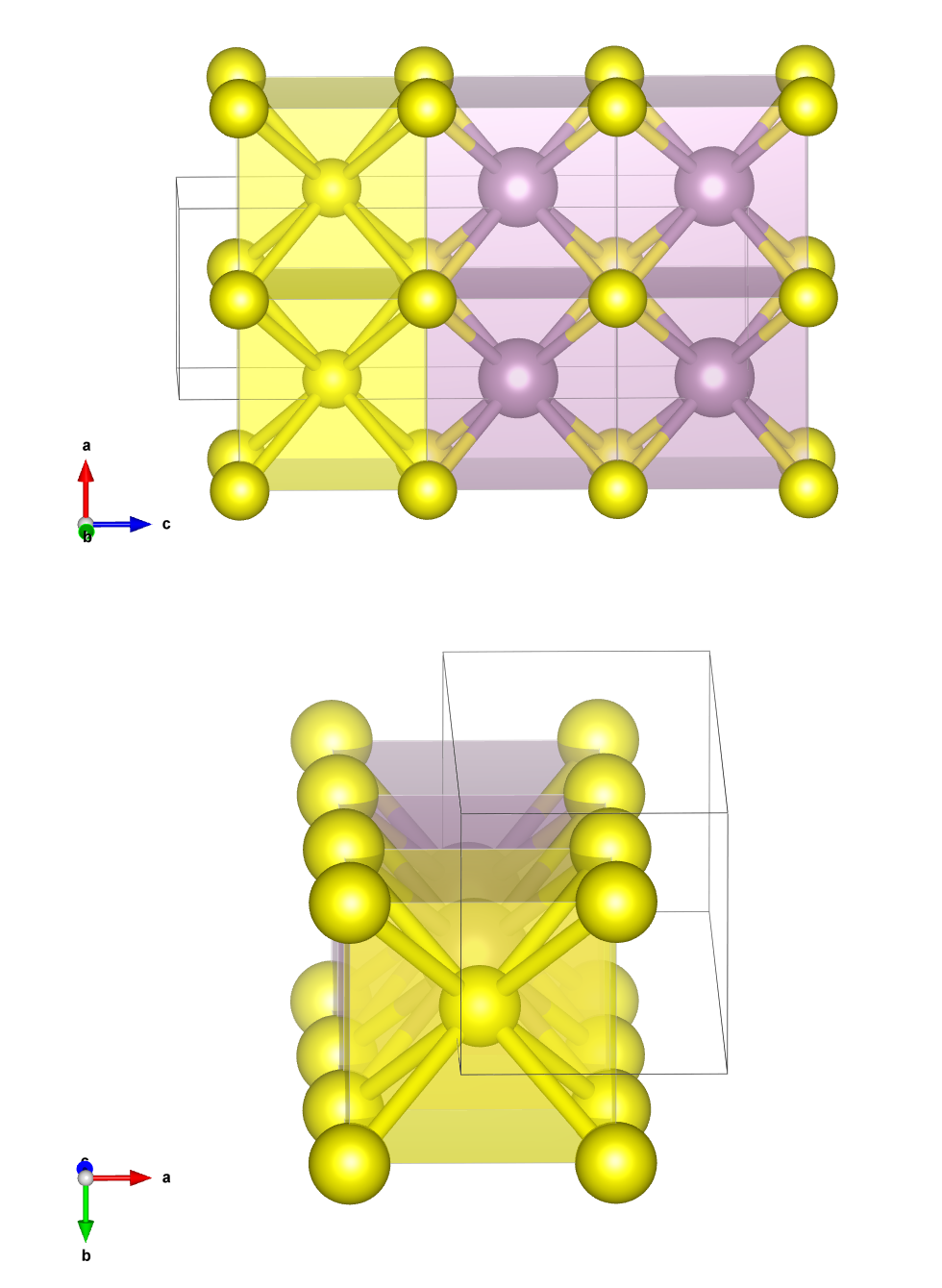}
\caption{Structure of the \textit{P4/mmm} phase. On the upper panel 2
  unit cells are shown.}
\label{fig:P4/mmm}
\end{figure}

\begin{figure}[h]
\centering
\includegraphics[width=0.95\textwidth]{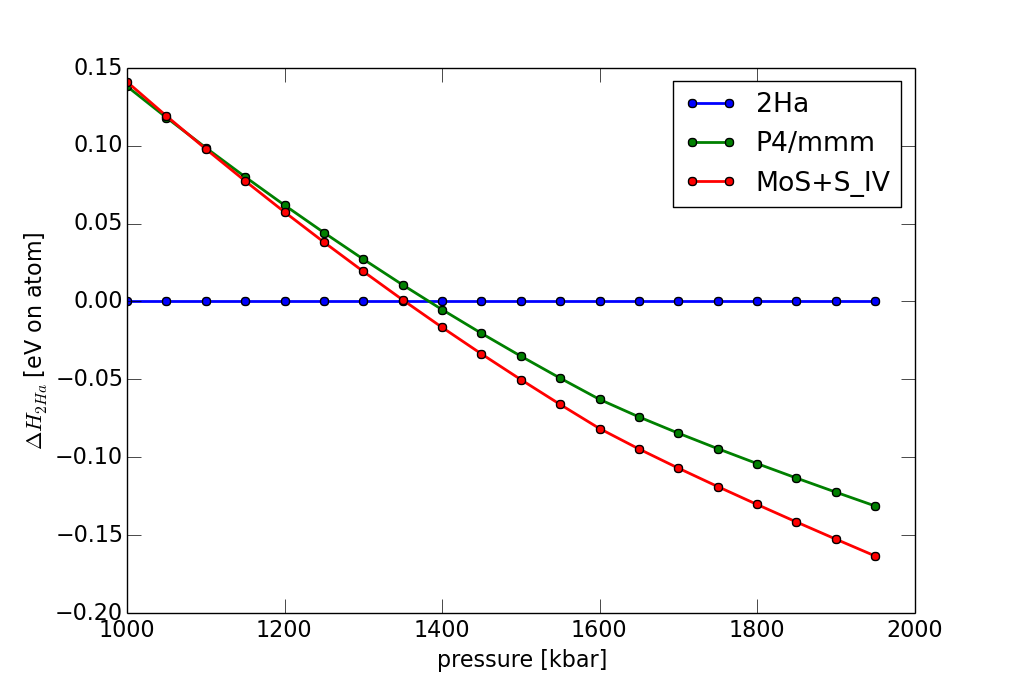}
\caption{Evolution of enthalpy with increasing pressure for various
  phases. The 2H$_a$ structure is taken as reference. Enthalpies were
  calculated with Quantum Espresso, using k-points mesh of
  $12\times12\times4$ for the 2H$_a$ and \textit{P4/mmm},
  $12\times12\times12$ for MoS and $13\times13\times13$ for SIV.}
\label{entcomp}
\end{figure}

\begin{figure}[h]
\includegraphics[width=0.95\textwidth]{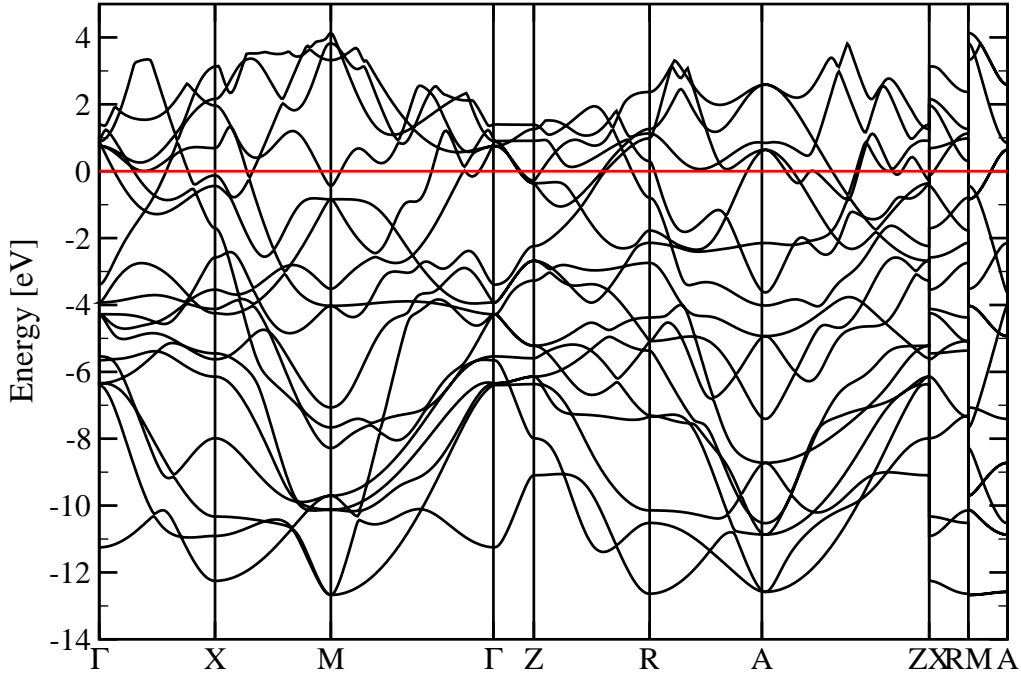}
\caption{Band structure for \textit{P4/mmm} calculated from QE at 150
  GPa. The Fermi level is shown with red line.}
\label{elstrct_P4MMM}
\end{figure}

\begin{figure}[h]
\includegraphics[width=0.95\textwidth]{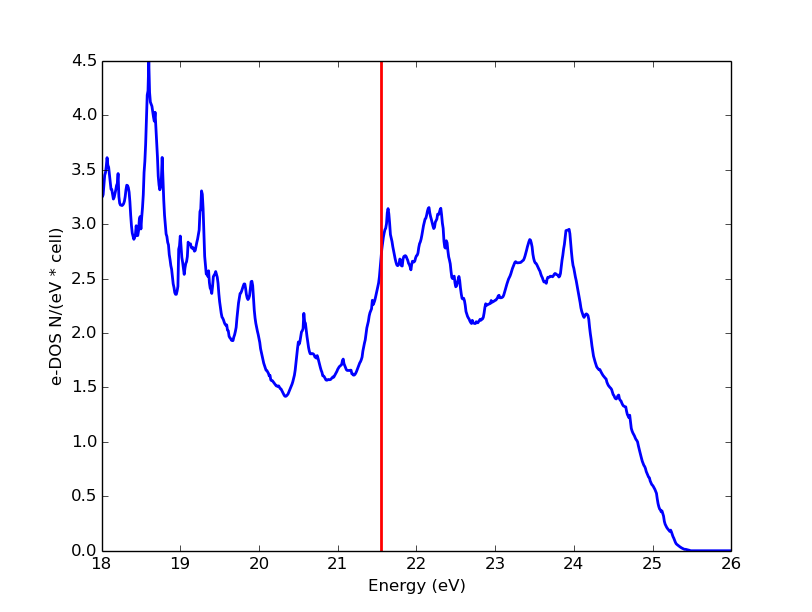}
\caption{e-DoS for the \textit{P4/mmm} structure calculated from QE at
  150 GPa. The Fermi level is shown with red line.}
\label{edos_P4MMM}
\end{figure}

\begin{figure}[h]
\centering
\includegraphics[width=0.95\textwidth]{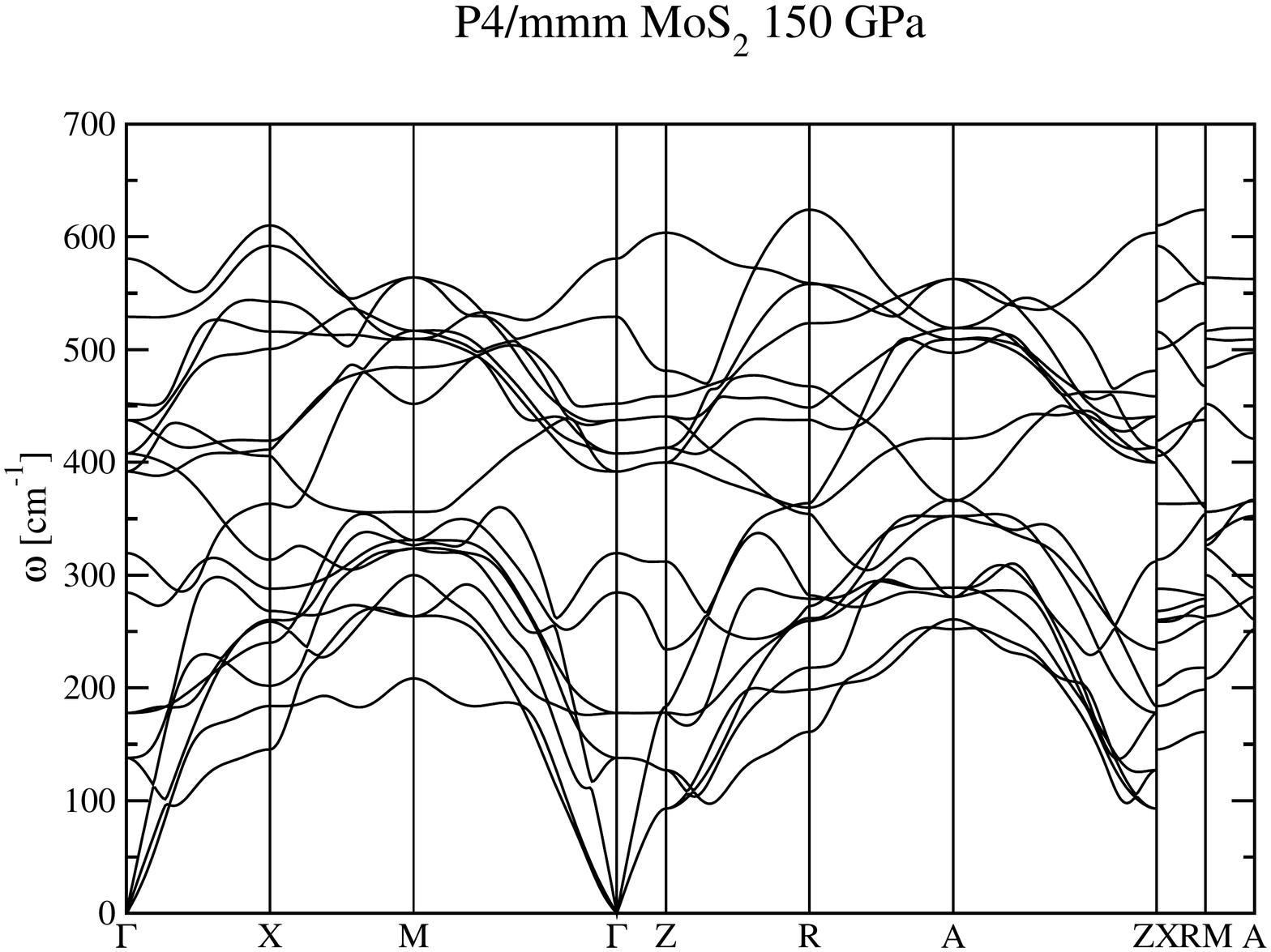}
\caption{Phonon dispersion in the \textit{P4/mmm} structure at 150
  GPa.}
\label{fig:phonons_P4/mmm_150GPa}
\end{figure}

Results of this extensive and unprejudiced high pressure structural
search recover and reproduce first of all the well-known 2H$_a$,
2H$_b$, 2H$_c$ and 3R layered structures. As was found in previous
work, 2H$_a$ has the lowest enthalpy among them in the selected
pressure range\cite{Hromadova2013}.  However, a new structure was also
found with 6 atoms/cell and space group \textit{P4/mmm}
(Fig.\ref{fig:P4/mmm}).  As shown by the calculated enthalpies in
Fig.\ref{entcomp} the tetragonal \textit{P4/mmm} competes strongly
with 2H$_a$ at sufficiently high pressure where, with a strongly first
order phase transition involving a volume drop of about 5 \% it
overcomes the 2H$_a$ structure at 138 GPa.  Fully relaxing and
optimizing the \textit{P4/mmm} structure at 150 GPa, we refined at
this pressure its structural and electronic properties.  The new
structure has a bimolecular cell like 2H$_a$, but unlike it is not
layered.  The \textit{P4/mmm} unit cell is tetragonal with lattice
parameters $a = 2.701 \; \mbox{\AA}, c =7.933 \; \mbox{\AA}$.  It
contains two Mo atoms, each surrounded by eight S atoms in a CsCl
local coordination, plus one S atom in an S-atom bcc cage. The band
structure and electronic density of states (Figs.~\ref{elstrct_P4MMM}
and \ref{edos_P4MMM}) show a strongly metallic character, as opposed
to the 2H$_a$ structure which even at this pressure is only weakly
metallic\cite{Hromadova2013}.  The calculated phonon dispersion
(Fig.\ref{fig:phonons_P4/mmm_150GPa}) which we obtained using QE shows
that at 150 GPa the \textit{P4/mmm} structure is locally stable. Based
on this electronic and phonon structure we calculate the
electron-phonon coupling parameter and find $\lambda \sim 0.75$, much
larger than that calculated for the 2H$_a$
structure\cite{Hromadova2013, Profeta}, suggesting a superconducting
temperature $T_c$ of about 15 K obtained assuming $\mu^*
=0.1$. Therefore a transformation $2H_a \to P4/mmm$ could in principle
lead to the abrupt onset of superconductivity at low temperatures and
high pressures.

Further calculations, however, suggest that a structural
transformation from $2H_a \to P4/mmm$ is not the likeliest equilibrium
occurrence in MoS$_2$ at high pressure.  The \textit{P4/mmm} structure
actually consists of adjacent nanosized elemental sulphur and MoS
crystals. That suggests another possible scenario, namely MoS + S
chemical decomposition.  Because MoS is not a well known phase,
additional calculations are needed in order to confirm or deny that
kind of high pressure transformation route.

We carried out a genetic algorithm search on MoS, and indeed found a
stable structure of MoS at 120 GPa, endowed precisely with the CsCl
structure, which is metallic.  The structure of elemental sulphur in
this pressure range is known to be the phase IV, body-centered
monoclinic with an incommensurate
modulation\cite{PhysRevB.71.214104}. Neglecting the modulation we
calculated the enthalpy curve corresponding to the MoS + S
decomposition enthalpy which is shown in Fig.\ref{entcomp}.  It should
be noted that inclusion of modulation could at most slightly decrease
the decomposition pressure, but will not increase it.  Thus,
decomposition with phase separation of 2H$_a \rightarrow$ MoS + S is
predicted to take place at or below 135 GPa, preempting the structural
transformation to the \textit{P4/mmm} phase. Decomposition of 2H$_a$
therefore represents the predicted thermodynamic evolution under
pressure. Since elemental sulphur in phase IV is known to superconduct
above 93 GPa\cite{PhysRevLett.99.155505} with $T_c > 10$ K, the
decomposition scenario should lead to the abrupt onset of
superconductivity in the S component.

Because the enthalpies of \textit{P4/mmm} and of MoS + S are so very
close, a structural transformation to the metastable \textit{P4/mmm}
phase, or some other similar phase with a larger unit cell, cannot be
ruled out and might take place experimentally, especially if
decomposition was protected by a high barrier, or simply if the
necessary diffusion of phase separated S turned out to be kinetically
too slow. Other possibilities might include intermediate routes such
as partial decomposition with formation of e.g., filamentary
sulphur. Alternatively, MoS$_2$ might conceivably survive in a
metastable 2H$_a$ structure even above 138 GPa, owing to the
presumably slow kinetics of experimental phase transformations and
separations.  Whether the predicted equilibrium transformations will
take place in ultra-high pressure MoS$_2$ and if so in what form will
depend exquisitely on these kinetic factors, and can only be
established experimentally.

\begin{acknowledgments}
  O.K. and R.M. were supported by the Slovak Research Development
  Agency under Contract APVV-0108-11 and R.M. was also supported by
  the contract APVV-0558-10. We also acknowledge support from the
  project implementation 26220220004 within the Research $\&$
  Development Operational Programme funded by the ERDF.  Part of the
  calculations was performed in the Computing Centre of the Slovak
  Academy of Sciences using the supercomputing infrastructure acquired
  in project ITMS 26230120002 and 26210120002 (Slovak infrastructure
  for high-performance computing) supported by the Research $\&$
  Development Operational Programme funded by the ERDF.  Work in
  Trieste was partly sponsored by EU-Japan Project LEMSUPER, by
  Sinergia Contract CRSII2$_1$36287/1, and by ERC Advanced Grant
  320796 - MODPHYSFRICT.  We are especially grateful to G. Profeta,
  A. Sanna for information on their unpublished calculations, and to
  Z. Chi, T. Kagayama, Y. Iwasa and K. Shimizu for discussion and
  information on experimental aspects.
\end{acknowledgments}

\end{document}